\documentclass[reprint,twocolumn,showpacs,nofootinbib]{revtex4}

\usepackage{mathbbol,upgreek}
\usepackage{graphicx}
\usepackage{color}
\usepackage[caption=false]{subfig}

\begin{document}

\title{Piezoelectrically-actuated time-averaged atomic microtraps}
\date{\today}
\author{A. D. West}
\email{adam.west@durham.ac.uk}
\affiliation{Department of Physics, Durham University, Rochester Building, South Road, Durham, DH1 3LE, UK.}
\author{C. G. Wade, K. J. Weatherill and I. G. Hughes}
\affiliation{Department of Physics, Durham University, Rochester Building, South Road, Durham, DH1 3LE, UK.}

\begin{abstract}             
We present a scheme for creating tight and adiabatic time-averaged atom-traps through the piezoelectric actuation of nanomagnetic structures. We show that potentials formed by the circular translation of magnetic structures have several advantages over conventional rotating-field techniques, particularly for high trap frequencies. As the magnitude of the actuation is changed the trapping potential can be changed adiabatically between harmonic 3D confinement and a toroidal trap.
\end{abstract}


\maketitle

The time-averaging technique is a well established method for trapping both charged and neutral particles \cite{paul90} via the creation of effective electromagnetic potentials which are not restricted by Maxwell's equations. The simplest static-field magnetic trap is given by a quadrupole \cite{migd85} and is now ubiquitous within ultracold physics. However, this geometry suffers from spin-flips which occur at the magnetic field zero \cite{berg97} and the associated loss rate becomes very high for tight traps. The problem of spin-flip losses can be overcome by creating a time-averaged potential (TAP) via the application of a rotating magnetic field \cite{petr95}, and resulted in the first observation of a Bose-Einstein condensate (BEC) \cite{BEC}. Since then stable static-field geometries with non-zero field minima have been demonstrated, such as the Ioffe-Pritchard trap \cite{erns97} or QUIC trap \cite{essl98}. However, the use of time-varying fields remains popular as it affords dynamic tuning of the potential and can produce more elaborate topologies, especially via the technique of RF-dressing \cite{mori06,lesa07,schu05,sher11}.

Microfabricated `atom chips' are becoming increasingly attractive due to the scalability and precision afforded by microfabrication techniques \cite{chipbook}. Power dissipation problems make further miniaturization difficult in devices based on current-carrying wires. However, patterned magnetic materials offer a promising route to chips with nanometric lengthscales \cite{allw06}. At these lengthscales the resulting potentials can exhibit extremely high trap frequencies and are expected to allow investigation of new physics \cite{folman}. However these benefits come at the cost of the presence of a zero in the magnetic field which leads to spin-flips that dramatically jeopardise the utility of the trap. Circumventing these losses in such tight traps using conventional techniques is extremely challenging, and often technically infeasible.

Here we propose an alternative approach to the creation of TAPs through the rapid mechanical oscillation of a magnetic field source by piezoelectric actuators. We show that one can make tight and adiabatic microtraps with significant advantages over convential TAPs, particularly at higher trap frequencies. We also find that these piezoelectrically-actuated TAPs (PATAPs) can be adiabatically transformed between spheroidal and toroidal geometry.

The utility of a TAP is characterised by two inequalities. Firstly we must achieve the time-averaging condition, i.e.\ the oscillation frequency, $\omega_{\rm TAP}$, must be higher than the characteristic trap frequency, $\omega_{\rm Trap}$. Secondly, to remain adiabatic we require that the minimum Larmor frequency, $\omega_{\rm L}$, is greater than $\omega_{\rm TAP}$. I.e.\ we require
\begin{equation}
\omega_{\rm L}>\omega_{\rm TAP}>\omega_{\rm Trap}.
\label{eq:ineq}
\end{equation}
We characterise this inequality by a parameter $\xi=\omega_{\rm L}/\omega_{\rm Trap}$. A useful TAP must also be sufficiently deep. The depth is defined by the minimum energy required to reach a position of the instantaneous zero point, and we label this $U_{\rm D}$.

In previous work we proposed the use of rotating magnetic fields to create tight atom traps based on fringing fields derived from nanomagnetic domain walls \cite{allw06,hayw11}. However, to produce a sufficiently adiabatic and deep trap one must apply rotating fields of several gauss. This is technically very challenging as it requires driving large currents through inductive coils at high frequencies (100~kHz-10~MHz).
\begin{figure}[!htb]
\centering
\includegraphics[width=\columnwidth]{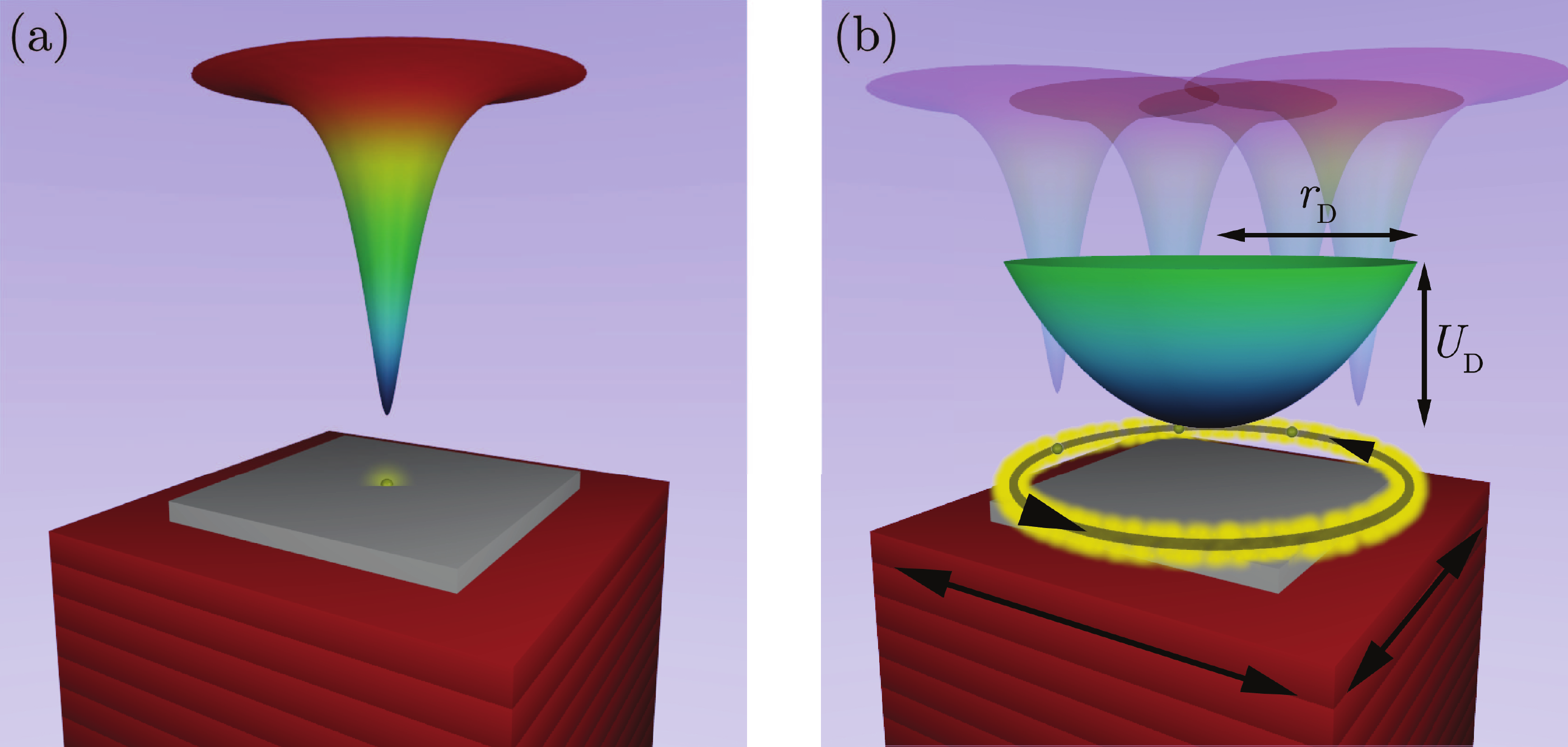}
\caption{(a) Schematic of the static quadrupole-like trapping potential resulting from a biased fringing field. (b) Schematic of the TAP resulting from circular actuation of the magnetic field source. $U_{\rm D}$ is the trap depth and $r_{\rm D}$ is the radius of oscillation. The TAP is truncated at $r_{\rm D}$ to illustrate the trap depth due to the instantaneous zero point. The shape of the potentials are quantitatively accurate but not to scale.}
\label{fig:piezoblend}
\end{figure}

By rapidly translating a magnetic field source we can achieve a TAP without the need for large, rapidly rotating fields. Consider a surface containing a static magnetic field source which produces fringing fields. For the purpose of calculations we assume this field source is a domain wall within a magnetic nanowire, but could be any point-like source. If the fringing fields emanate in the $+z$ direction, applying a static bias field in the $-z$ direction produces a static magnetic field, $\vec{B}_{\rm S}(\vec{r})$, which approximates a 3D quadrupole near to the zero point \cite{allw06}. We define the zero point to be centred on the origin. As the magnitude of the bias field increases, the trap is formed closer to the magnetic field source, and the field gradient increases. We characterise the static field by the field gradient in directions perpendicular to the bias field, $B^{\prime xy}_{\rm S}$ \footnote{We note that by symmetry $B^{\prime x}_{\rm S}=B^{\prime y}_{\rm S}=-B^{\prime z}_{\rm S}/2$.}. If we translate the surface in a circle of radius $r_{\rm D}$ at frequency $\omega_{\rm TAP}$ then the time-dependent position is given by
\begin{equation}
\vec{r}(t)=r_{\rm D}\cos(\omega_{\rm TAP}t)\hat{x}+r_{\rm D}\sin(\omega_{\rm TAP}t)\hat{y}.
\label{eq:ppos}
\end{equation}
This circular motion replicates the `circle of death' produced when using time-orbiting fields \cite{petr95}.

For sufficiently high $\omega_{\rm TAP}$ the trapped atoms will experience an effective potential, $U$, given by the time average of the magnetic field magnitude:
\begin{equation}
\label{eq:u}
U=m_Fg_F\mu_{\rm B}\frac{\omega_{\rm TAP}}{2\pi}\int_0^{2\pi/\omega_{\rm TAP}}\left|\vec{B}_{\rm S}(\vec{r}(t))\right|{\rm d}t.
\end{equation}
This time-averaging procedure is shown schematically in Fig.~\ref{fig:piezoblend}. The effect is that the bottom of the trapping potential becomes harmonic with a finite field minimum.

\begin{figure}[!htb]
\centering
\includegraphics[width=8.4cm]{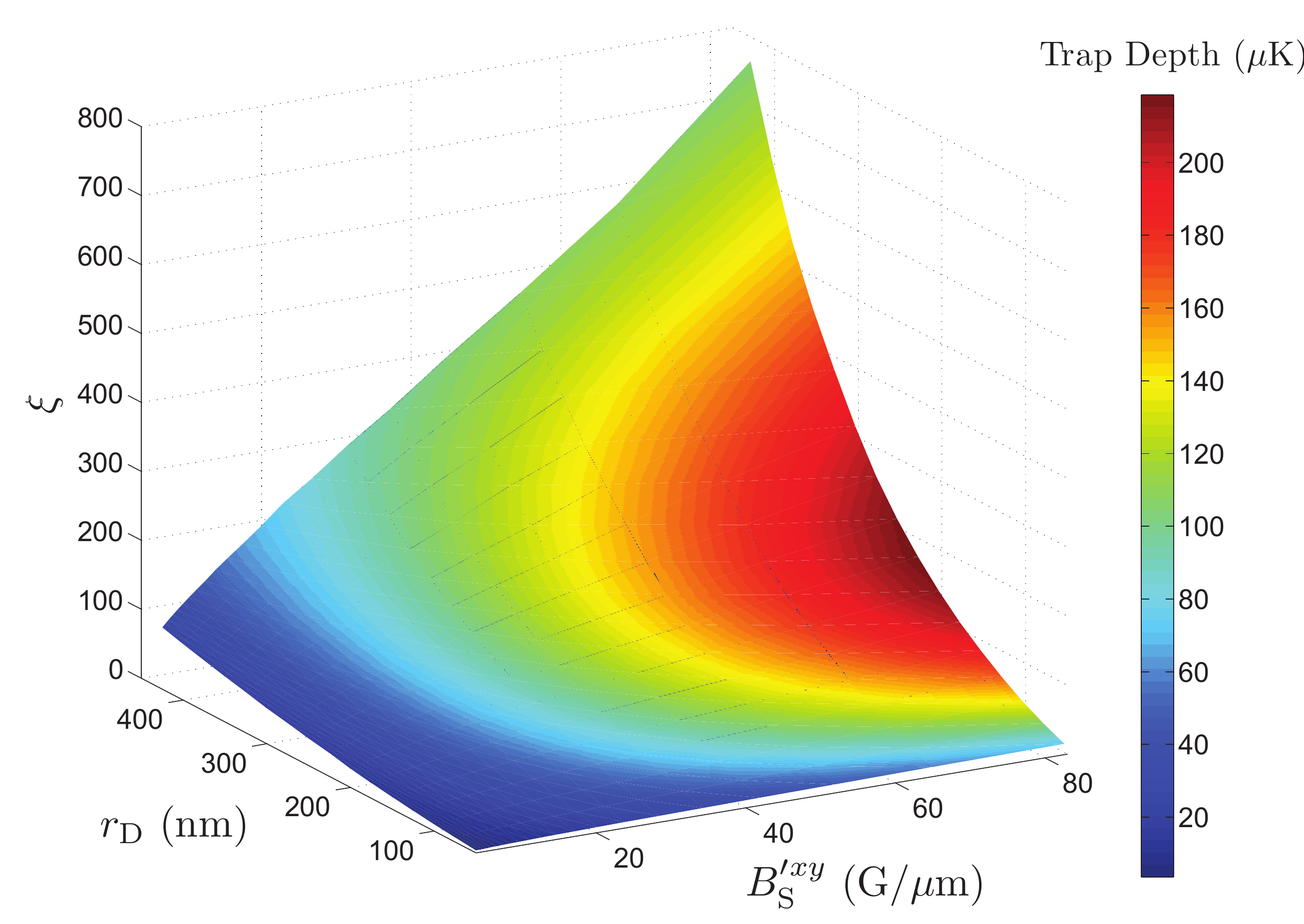}
\caption{Variation of PATAP characteristics with the amplitude of motion and the tightness of the static potential. $B_{\rm S}^{\prime xy}$ is the gradient of the static magnetic field in the $x$ and $y$ directions, $r_{\rm D}$ is the amplitude of movement and $\xi$ is the adiabaticity parameter equal to $\omega_{\rm L}/\omega_{\rm Trap}$. Deep and adiabatic traps are produced at $r_{\rm D}$ of a few hundred nanometers.}
\label{fig:patapadiab}
\end{figure}

To gauge the utility of the PATAP scheme we consider the figures of merit $\xi$ and $U_{\rm D}$ of traps containing $^{87}$Rb atoms, for a range of $r_{\rm D}$ and $B_{\rm S}^{\prime xy}$, via numerical calculation of the time-averaged potential. The static field was calculated according to an analytic model for nanomagnetic fringing fields \cite{west12}. The resulting data are shown in Fig.~\ref{fig:patapadiab}.

The first conclusion to draw from the data is that it is possible to produce deep and adiabatic traps for displacements of only a few hundred nanometres. There are some clear trends in the data. Perhaps the more notable is that for a given $r_{\rm D}$ the trap becomes deeper and more adiabatic (larger $\xi$) for higher static field gradients, i.e.\ the technique works better for tighter traps. This is intuitive as for a given $r_{\rm D}$ a higher field gradient results in a larger change in magnetic field and hence a larger $U_{\rm D}$ and $\omega_{\rm L}$. This trend is in direct contrast to conventional TAPs: for a given amplitude of rotating field the adiabaticity decreases for larger static field gradients \cite{hayw11}. Again this is intuitive -- for larger field gradients a larger rotating field is required to produce a given movement of the instantaneous zero point.

We also observe that $\xi$ increases as $r_{\rm D}$ increases; larger movement increases $\omega_{\rm L}$ whilst also reducing $\omega_{\rm Trap}$. This behaviour is also observed for TAPs using rotating fields -- as the size of these fields increases, $\xi$ increases. For the deepest trap shown in Fig.~\ref{fig:patapadiab}, obtained with $B_{\rm S}^{\prime xy}\approx 80~$G/$\upmu$m and $r_{\rm D}=250$~nm, $\omega_{\rm Trap}=2\pi~\times~127$~kHz. For the point with largest $\xi$, $\omega_{\rm Trap}=2\pi~\times~54$~kHz.

The final trend shown in Fig.~\ref{fig:patapadiab} is that there is a strong dependence of $U_{\rm D}$ on $r_{\rm D}$. Increasing the amplitude of oscillation initially causes an increase in the depth of the trap. For larger $r_{\rm D}$ the time-averaged potential is larger at the position of the instantaneous zero point. However, further increase in $r_{\rm D}$ causes a subsequent decrease in the trap depth. This is due to the finite size of the static potential, illustrated in Fig.~\ref{fig:piezoblend}. As $r_{\rm D}$ becomes larger than this characteristic size, the shape of the PATAP begins to change -- there is less `overlap' at the centre of the PATAP, and it begins to flatten out such that the trap depth is reduced. 

\begin{figure}[!htb]
\centering
\subfloat{\includegraphics[width=4.2cm]{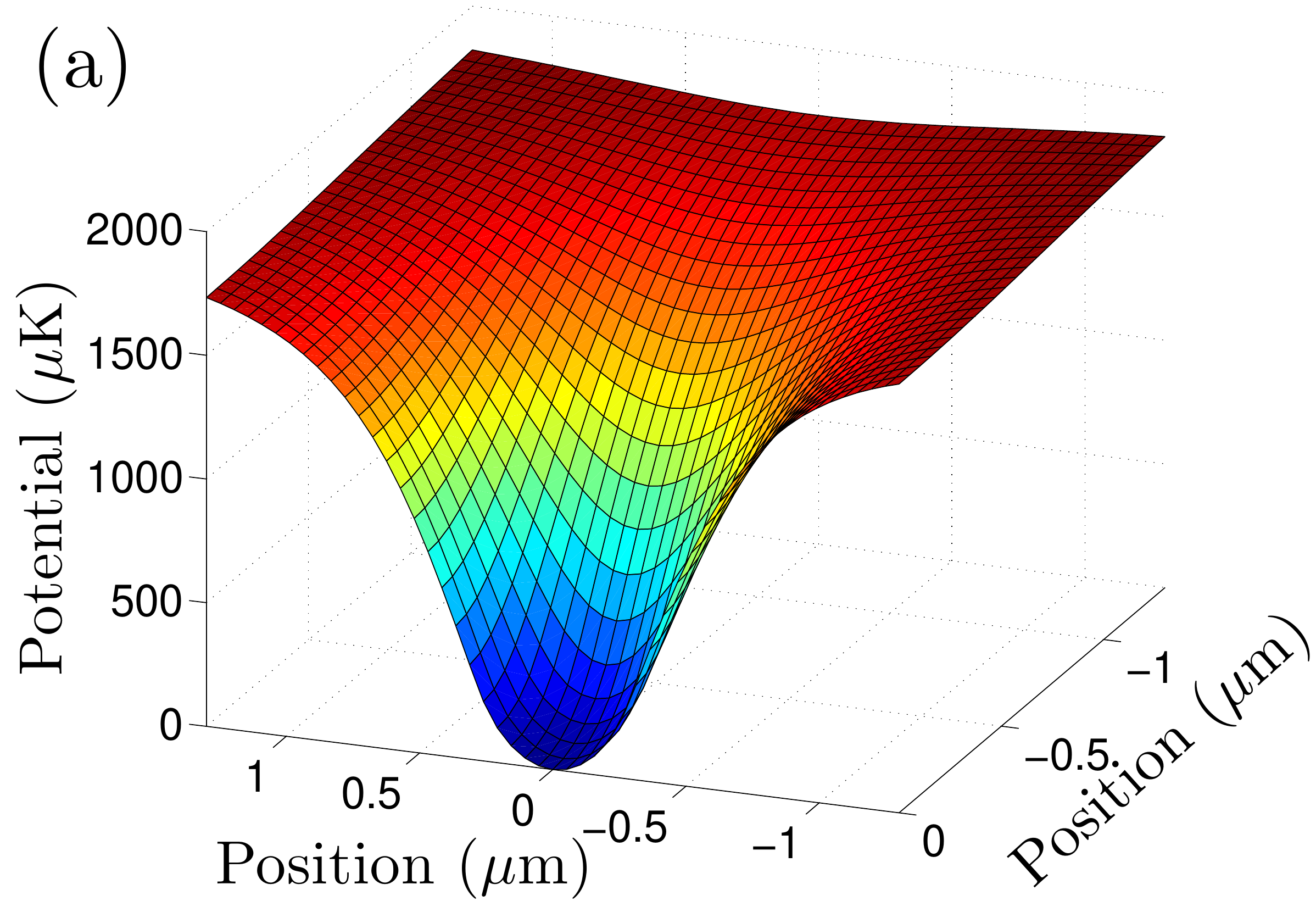}}
\subfloat{\includegraphics[width=4.2cm]{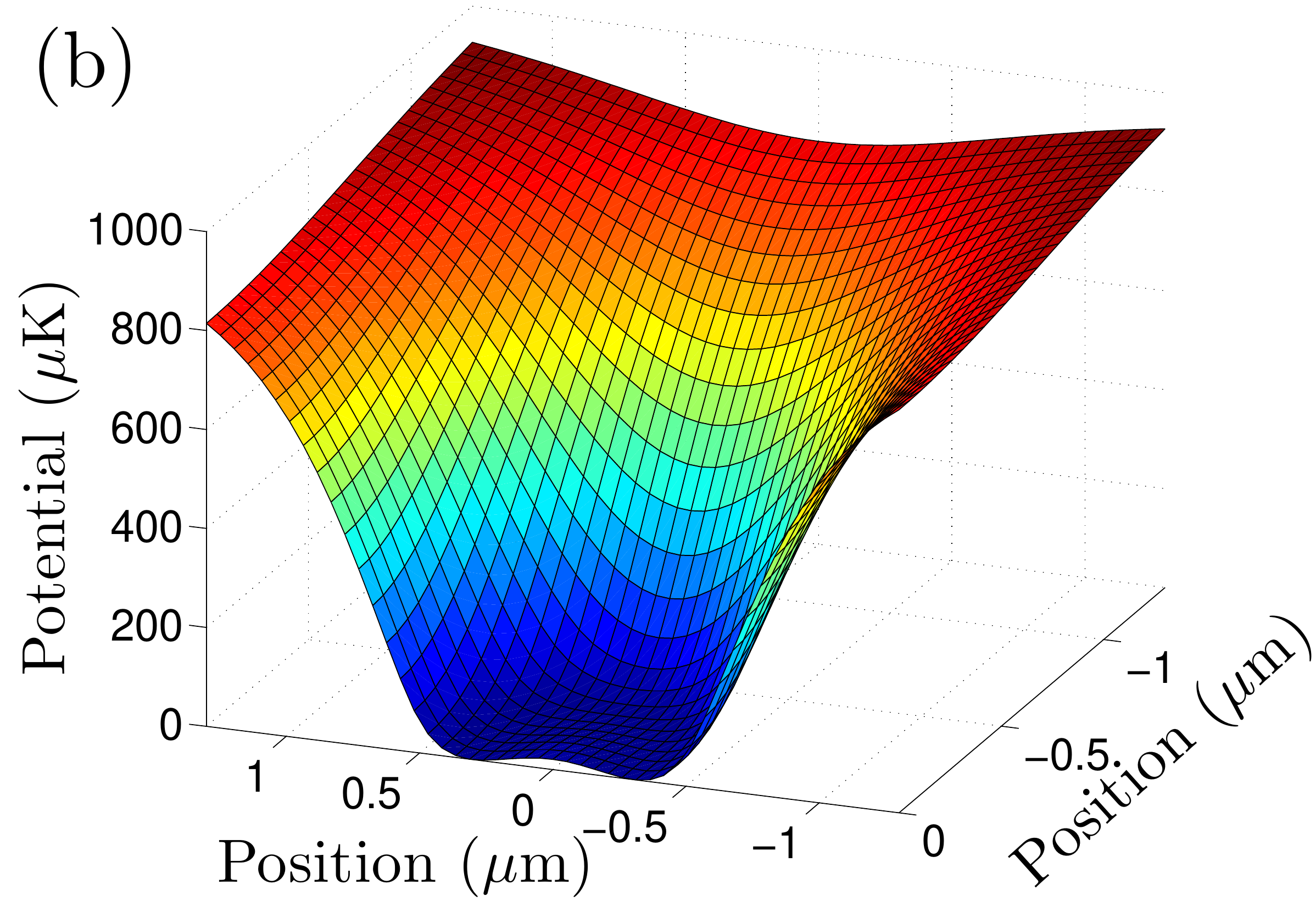}}\\
\subfloat{\includegraphics[width=4.2cm]{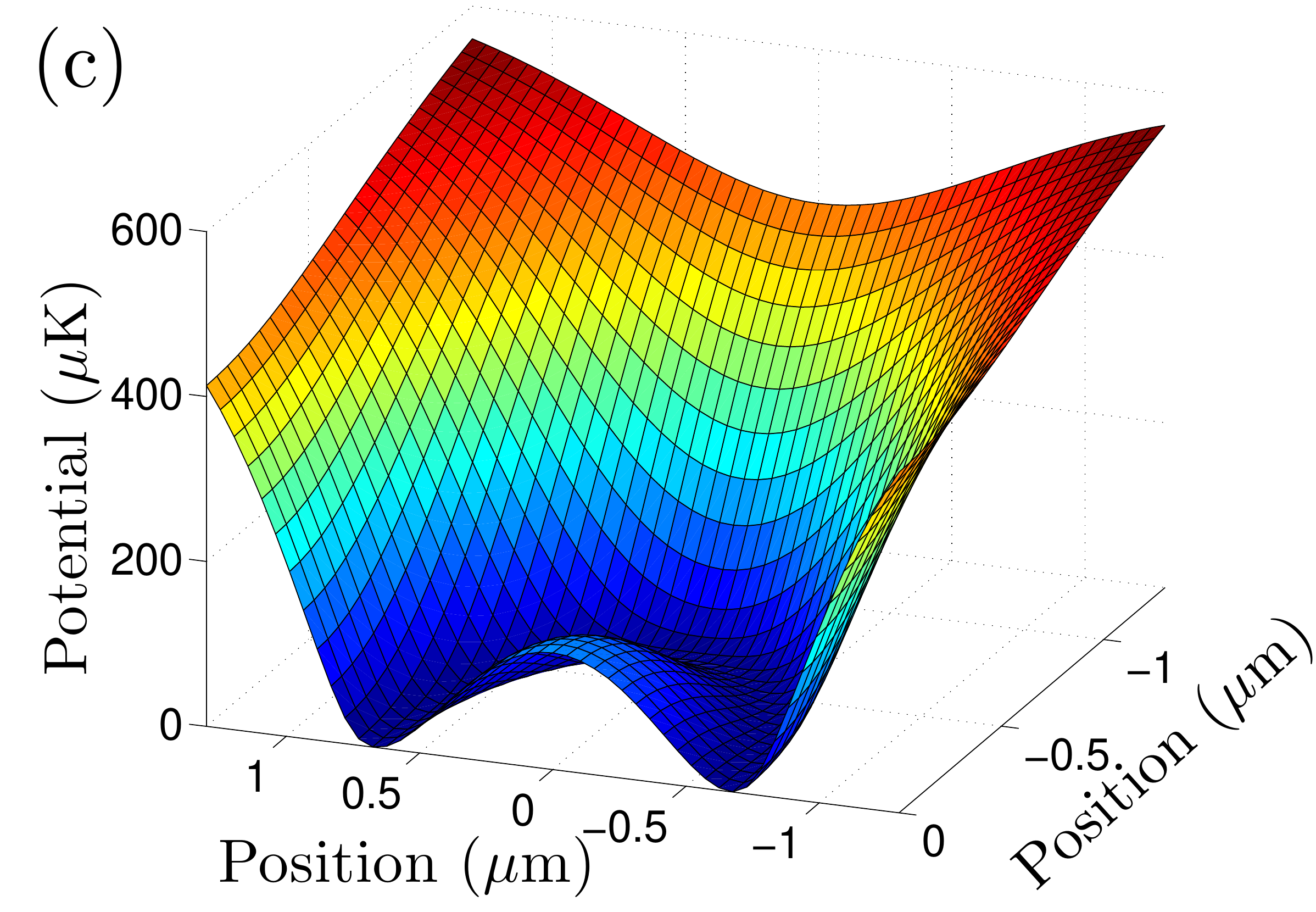}}
\subfloat{\includegraphics[width=4.2cm]{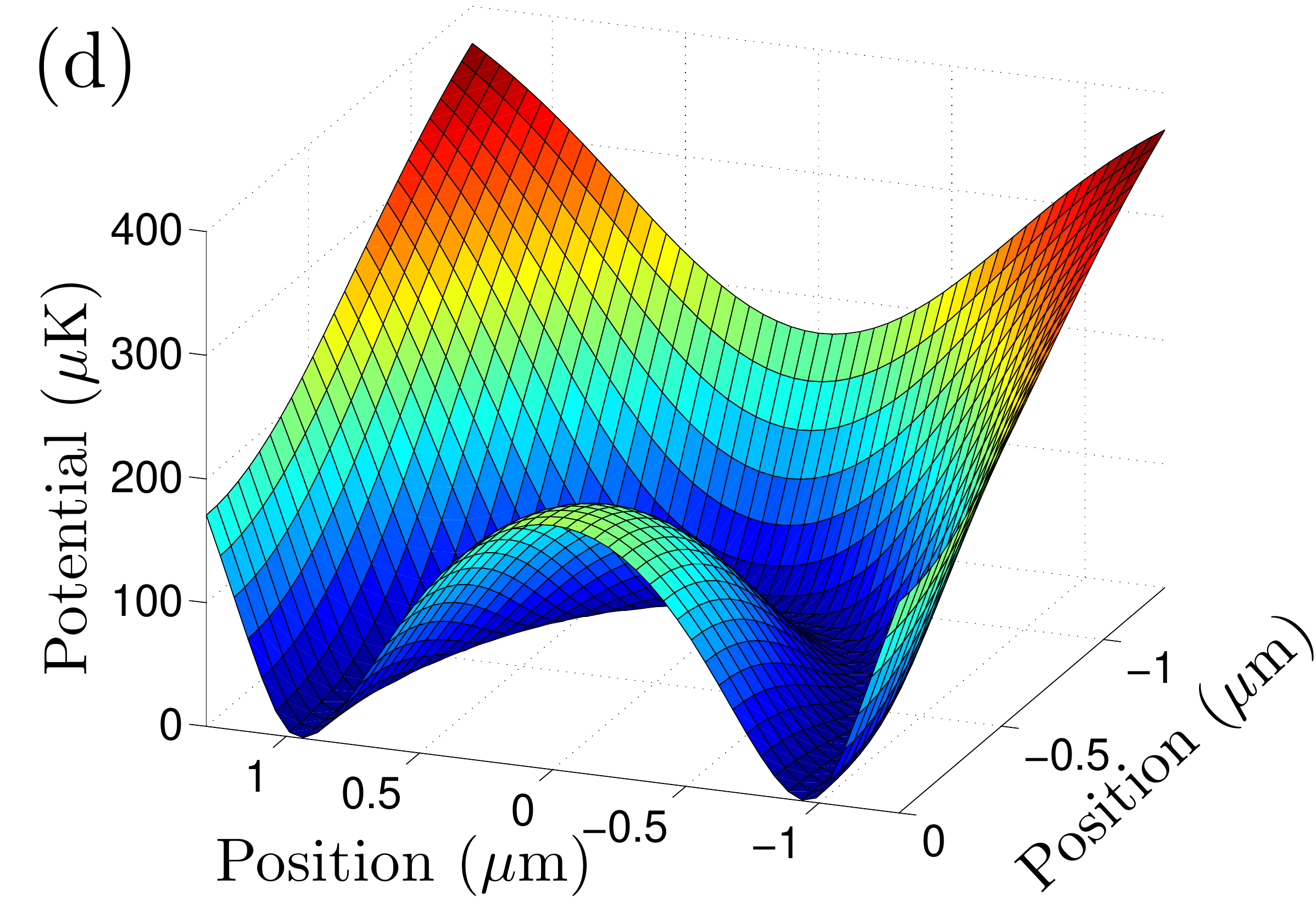}}
\caption{The resulting time-averaged potential for a piezoelectrically-actuated nanomagnetic domain wall for a range of oscillation radii, $R$. (a)-(d) correspond to $R=0.25~\upmu$m, 0.50~$\upmu$m, 0.75~$\upmu$m and 1.00~$\upmu$m respectively. The character of the trapping geometry changes from simple 3D confinement to a ring trap with 1D confinement as $R$ increases.}
\label{fig:pataptrans}
\end{figure}

Continuing to increase $r_{\rm D}$ leads to a change in the topology of the trap. For sufficiently large $r_{\rm D}$, a toroidal trapping potential is formed. The transition in the trap shape is shown in Fig.~\ref{fig:pataptrans}.
The position of the locus of trap minima is on a ring slightly smaller than the circle describing the path of the instantaneous zero point. As $r_{\rm D}\rightarrow\infty$ the locations of the time-averaged minimum and the instantaneous minimum coincide. This picture is very similar to the tailored optical potentials that have been realised through the modulation of an optical dipole trap \cite{hend09,zimm11}. It is worth noting that the production of these types of more complex potentials is not possible for a conventional TAP -- the time-averaging procedure must be combined with another technique such as RF-dressing \cite{sher11} or an additional optical potential \cite{heat08}. Alternative methods such as an inductively coupled ring trap have also been proposed \cite{grif08}.

\begin{figure}[!t]
\centering
\vspace{-15pt}
\subfloat{\includegraphics[width=8.2cm]{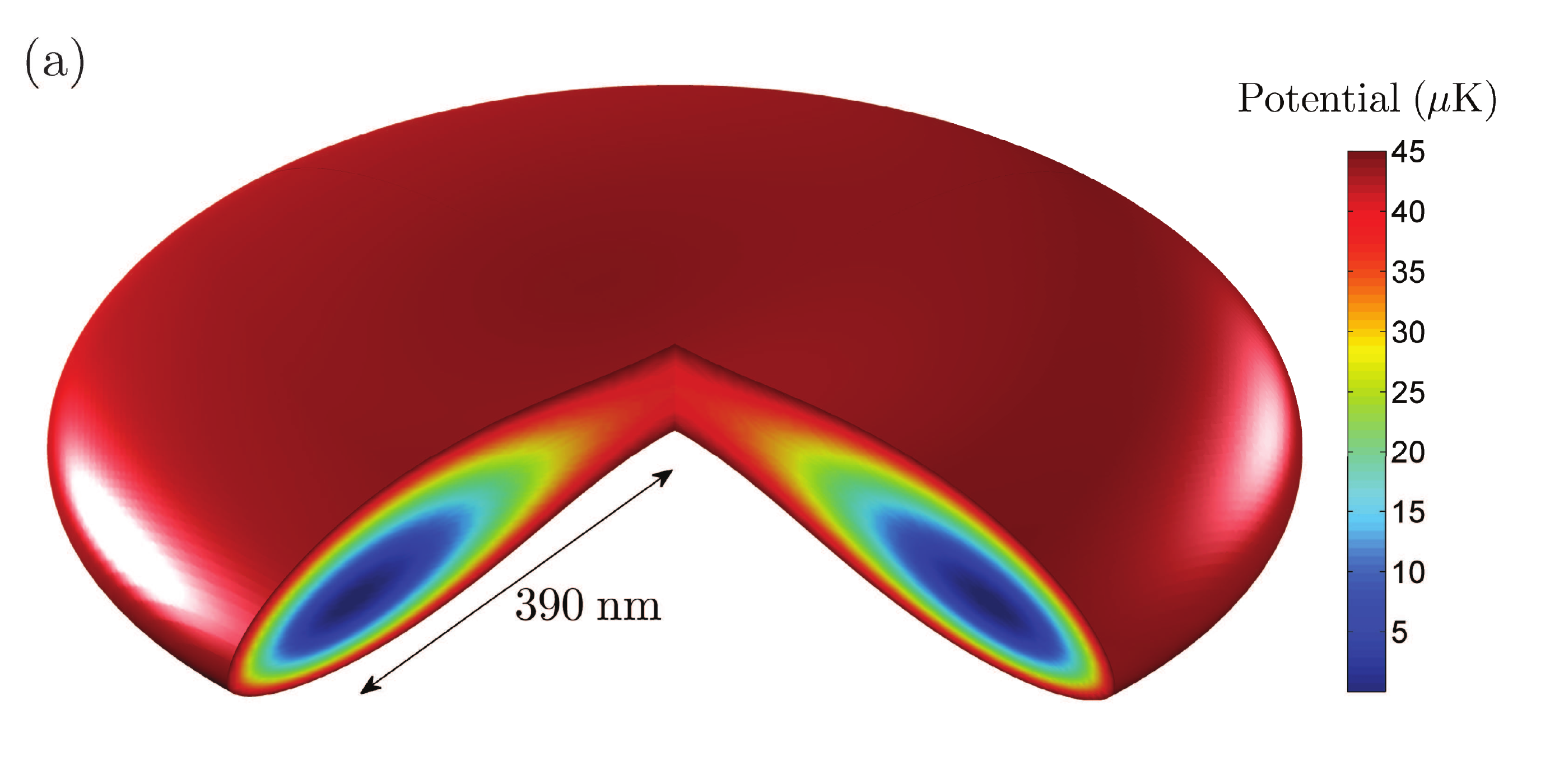}}\\
\vspace{-10pt}
\subfloat{\includegraphics[width=4.2cm]{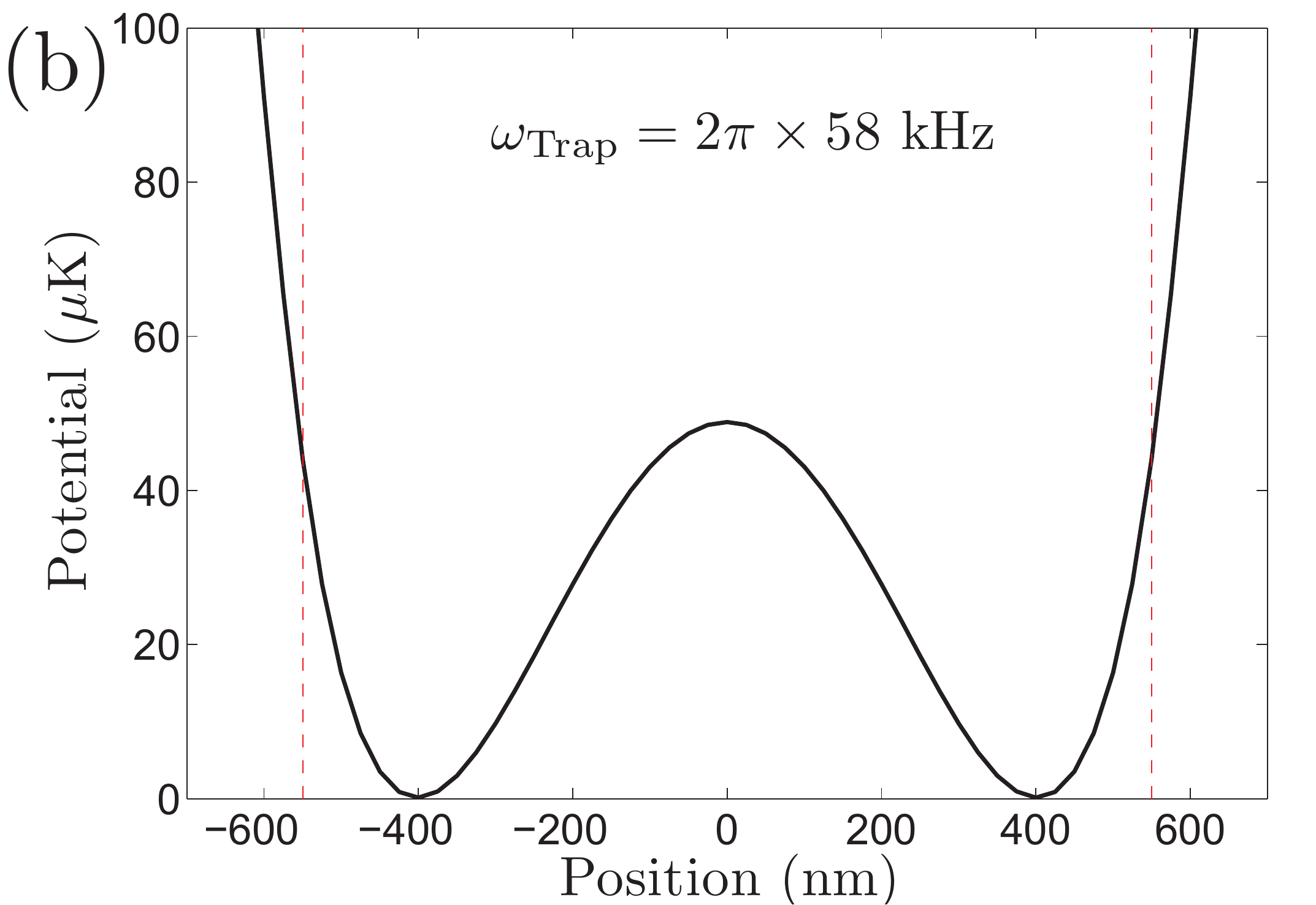}}
\subfloat{\includegraphics[width=4.2cm]{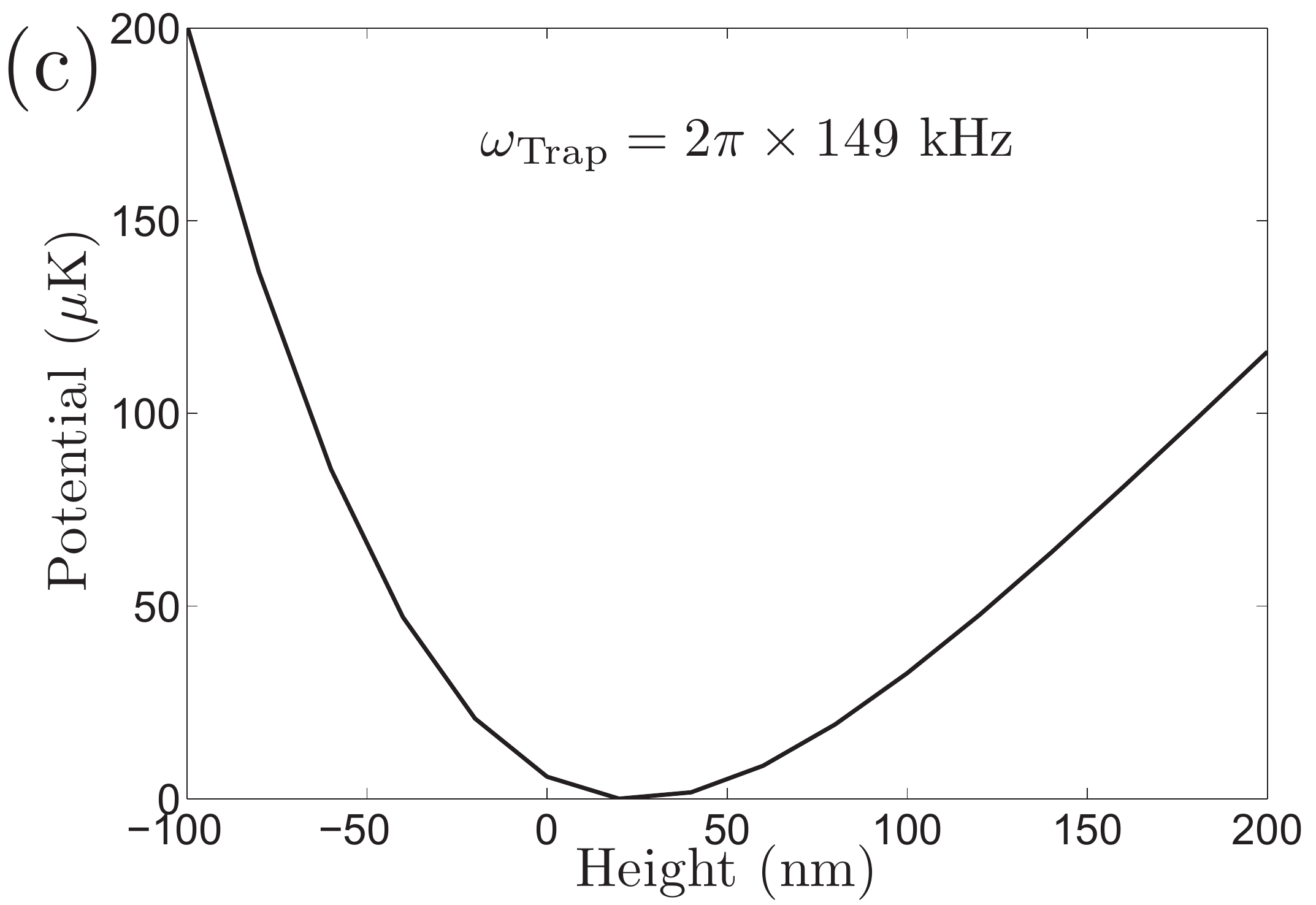}}
\caption{A toroidal trap formed using the PATAP scheme. (a) shows the general shape of the potential illustrated by an isosurface at $U_{\rm D}=45~\upmu$K. (b) and (c) show slices through the trap minimum in the radial and vertical directions respectively. The height is quoted relative to that of the instantaneous zero point. In (b), red dashed lines indicate the position of the instantaneous zero point. The trap is formed via oscillation of amplitude 550~nm. $\xi$ is given by $\omega_{\rm L}/\omega_{\rm Trap}=2\pi~\times~48$~MHz / $2\pi~\times~75~$kHz $\approx 600$.}
\label{fig:ring}
\end{figure}

With the change in topology, an additional definition of trap depth is introduced -- that defined by the central barrier of the torus. The smaller of the two definitions of trap depth is quoted. It is now a more difficult task to find suitable parameters to produce an adiabatic and deep trapping potential. An example is illustrated in Fig.~\ref{fig:ring}.

We observe that the resulting potential is again very tight and adiabatic. The minimum Larmor frequency is $2\pi~\times~48$~MHz and the trap frequency is $2\pi~\times~75~$kHz, yielding a value of $\xi$ of around 600. The trap depth is limited by the trajectory of the instantaneous zero point and has a value of 45~$\upmu$K. The height of the central barrier is 49~$\upmu$K. Thus we expect this ring potential to be ideally suited for tightly confining ultracold atoms in a toroidal geometry.

Because the same type of actuation can be used to create both a spheroidal and a toroidal geometry, we can adiabatically evolve the shape of the potential by ramping the amplitude of oscillation. Atoms initially loaded into a PATAP with moderate amplitude of oscillation could then be transferred into a ring trap whilst undergoing forced evaporation. The trajectory of the instantaneous zero point would describe a spiral and the potential would evolve as illustrated in Fig.~\ref{fig:pataptrans}.

We will now briefly discuss the experimental feasibility of the PATAP scheme. There are many vacuum compatible dual-axis commercial piezoelectric actuators which exhibit resonant motion in the $\sim$100~kHz--1~MHz frequency range with displacements of up to several microns, see for example \cite{pi}. A significant advantage of using a piezoelectric device is that it presents a capacitive load which draws very little current. This makes driving such a device much easier than producing rotating magnetic fields at high frequencies. The PATAP method is also less intrusive as it circumvents the need for large coils/antennae within a cold-atom setup. The field of piezoelectric microelectromechanical systems (MEMS) is already well developed and has demonstrated the ease with which one can amalgamate piezoelectric material with lithographically produced objects \cite{mems}. There are also devices which have combined piezoelectric actuators with nanomagnetic material \cite{danpiezo}, a feature that clearly lends itself to applications with atom-chip experiments. As with other tight magnetic traps \cite{fortzimm}, careful compression and spatial mode-matching will be required to load a PATAP from a magneto-optical trap.

We have presented a novel method for producing time-averaged potentials (TAPs). The resulting potentials have very high trap frequencies, whilst remaining adiabatic and deep. The small scale of these devices shows promise in their ability to be incorporated into atom chip type applications. We have also demonstrated that this piezoelectrically-actuated TAP scheme is versatile and could be used to produce a toroidal atom trap directly from a simple magnetostatic field.
 
We would like to thank Ben Sherlock, Barry Garraway and Simon Cornish for useful discussions and the UK Engineering and Physical Sciences Research Council for funding under grant EP/F025459/1.

\end{document}